\documentclass[12pt]{article}

\newcommand{\blind}{1}

\usepackage{authblk}
\usepackage{amsmath, amsthm, amssymb, bm}
\usepackage{graphicx, psfrag, epsf, subcaption}
\usepackage{enumerate}
\usepackage{natbib}
\usepackage{textcomp}
\usepackage[hyphens]{url} % not crucial - just used below for the URL
\usepackage{verbatim}

\usepackage{subfiles} %%%%%%%%%%%%%%%%%%%%%%%%%%% TEST
\usepackage{dsfont}
\usepackage{xcolor} % color text
\usepackage{hyperref}
\hypersetup{%
	colorlinks = false
}

\usepackage{verbatim}

\usepackage[utf8]{inputenc}

\usepackage{float} % For placing the figure

\usepackage{listings}
\usepackage{adjustbox}
\usepackage{multirow} % multiline of table

\usepackage{lineno} % add line number

\usepackage{moreverb}
\usepackage{amsmath}
\usepackage{amsthm}
\usepackage{array}
\newcolumntype{P}[1]{>{\centering\arraybackslash}p{#1}}
\newcolumntype{M}[1]{>{\centering\arraybackslash}m{#1}}
\usepackage{dsfont}
\usepackage{threeparttable}
\usepackage{booktabs}
\usepackage{rotating}
\usepackage{hyperref}
\hypersetup{hidelinks}
\usepackage{citeall}
\usepackage{float}
\usepackage{mathtools}
\usepackage{xcolor}
\usepackage[title]{appendix}

\usepackage{lineno}
\newcommand\BibTeX{{\rmfamily B\kern-.05em \textsc{i\kern-.025em b}\kern-.08em
		T\kern-.1667em\lower.7ex\hbox{E}\kern-.125emX}}

\usepackage[ruled,vlined]{algorithm2e}

\begin{document}
	
	\def\spacingset#1{\renewcommand{\baselinestretch}%
		{#1}\small\normalsize} \spacingset{1}

	%%%%%%%%%%%%%%%%%%%%%%%%%%%%%%%%%%%%%%%%%%%%%%%%%%%%%%%%%%%%%%%%%%%%%%%%%%%%%%
	\if1\blind
	{
		\title{\bf Q-learning in Dynamic Treatment Regimes with Misclassified Binary Outcome}
		
		\author[1]{Dan Liu\thanks{Corresponding author, email: dliu372@uwo.ca}}
		\author[1,2]{Wenqing He}
		\affil[1]{Department of Statistical and Actuarial Sciences, University of Western Ontario, London, Ontario, Canada N6A 5B7}
		\affil[2]{Department of Oncology, University of Western Ontario, London, Ontario, Canada N6A 5W9}
		\date{}                     %% if you don't need date to appear
		\setcounter{Maxaffil}{0}
		\renewcommand\Affilfont{\itshape\small}
		\maketitle
	} \fi
	
	\if0\blind
	{
		\title{ \bf Q-learning in Dynamic Treatment Regimes with Misclassified Binary Outcome}
		
		\author[1]{  }
		%\date{}
		\setcounter{Maxaffil}{0}
		\renewcommand\Affilfont{\itshape\small}
		\maketitle
	} \fi
	
	\begin{abstract}
		The study of precision medicine involves dynamic treatment regimes (DTRs), which are sequences of treatment decision rules recommended by taking patient-level information as input. The primary goal of the DTR study is to identify an optimal DTR, a sequence of treatment decision rules that leads to the best expected clinical outcome. Statistical methods have been developed in recent years to estimate an optimal DTR, including Q-learning, a regression-based method in the DTR literature. Although there are many studies concerning Q-learning, little attention has been given in the presence of noisy data, such as misclassified outcomes. In this paper, we investigate the effect of outcome misclassification on Q-learning and propose a correction method to accommodate the misclassification effect. Simulation studies are conducted to demonstrate the satisfactory performance of the proposed method. We illustrate the proposed method in two examples from the National Health and Nutrition Examination Survey Data I Epidemiologic Follow-up Study and the smoking cessation program.
	\end{abstract}
	
	\noindent%
	{\it Keywords:} Dynamic treatment regimes, outcome misclassification, precision medicine, Q-learning, validation data
	
	\spacingset{1.45} % DON'T change the spacing!
	
	%\linenumbers
	
	\section{Introduction}\label{sec-1}

	Precision medicine presents a new strategy in healthcare in which the treatment is adapted to each patient based on the patient-specific characteristics. It considers patients' heterogeneity and provides a dynamic personalized treatment strategy. Dynamic treatment regimes (DTRs) are the sequences of treatment decision rules to a patient by taking the patient's characteristics and treatment history into account\citep{chakraborty2013statistical}. The main objective of the DTR framework is to determine an optimal DTR, a sequence of treatment decision rules that leads to the best long-term clinical outcome.   
	
	There is substantial literature on the statistical methods to estimate an optimal DTR in various contexts \citep{watkins1989learning, murphy2003optimal, robins2004optimal, van2007causal, robins2008estimation, henderson2010regret, goldberg2012q, schulte2014q, huang2014optimization, wallace2015doubly, simoneau2020estimating}, among which the DTR with discrete-valued outcomes is rarely considered. However, discrete outcomes popularly arise in medical applications. For example, in the smoking cessation studies reported in \cite{lee2013effectiveness}, one important clinical objective is to identify a dynamic treatment regime that helps smokers quit smoking, where whether or not quitting smoking is the outcome of interest. To accommodate discrete-valued outcomes in DTR, Q-learning was extended to binary outcomes and count outcomes \citep{moodie2014q}, and a likelihood-based approach was proposed to estimate the dynamic treatment regimes with ordinal outcomes \citep{ghosh2018comparison}. Bayesian approaches were also developed to estimate the optimal DTR with binary outcomes \citep{artman2020bayesian}.  
	
	Although the existing methods are intuitive and useful in many aspects, the validity of the methods hinges on the assumption that the variables  are precisely measured. However, in practice, this assumption is often violated. In smoking cessation studies, for example, the self-reported smoking status involves misclassification. In a smoking cessation program at St. Joseph's Hospital, some smokers were observed to misreport their smoking status \citep{lee2013effectiveness}. In another example, the National Health and Nutrition Examination Survey Data I Epidemiologic Follow-up Study (NHEFS) focusing on the smoking cessation status from the cigarette smokers, the collected smoking status is also subject to misclassification. It has been documented that ignoring the misclassification in the response may yield misleading analysis results \citep{carroll2006measurement, yi2017statistical}. Therefore, obtaining an optimal DTR by using the misclassified outcome may cause erroneous results. Motivated by these examples, we aim to address the outcome misclassification effect in the estimation of dynamic treatment regimes. 
	
	Methods have been developed in the literature to correct the misclassification in a binary response. Maximum likelihood estimation (MLE) was utilized to correct for the misclassified outcome with a modified relationship between true and observed outcomes \citep{hausman1998misclassification, neuhaus1999bias, lyles2010sensitivity}, and it was further used for case-control studies in a validation/main data context \citep{lyles2011validation}. Semiparametric methods have been developed as an alternative to the MLE method to handle the misclassified response \citep{pepe1992inference, yi2017statistical}. EM algorithm was proposed to correct the misclassification \citep{magder1997logistic}. A multiple imputation approach was introduced to deal with misclassified outcomes based on the validation data subsample. 
	
	Although there are many studies on dynamic treatment regimes and misclassified outcomes separately, to the best of our knowledge, there is no research work on considering both issues together. In this paper, we study the misclassification effect of the binary outcomes on DTR via Q-learning. Internal validation data are assumed to be available, in which the true and misclassified outcomes are both observed. The maximum likelihood estimation is employed as a correction method to accommodate the misclassification effect on optimal DTR selection using Q-learning. 
	
	The rest of the article is organized as follows. In Section \ref{qlearn}, the Q-learning with binary outcomes is presented. The misclassification process for the binary outcome is introduced in Section \ref{qlearnMis}. The correction method to account for the misclassification effect on optimal DTR via Q-learning is described in Section \ref{method}. In Section \ref{sim}, we conduct simulation studies to evaluate the performance of the proposed method in one-stage and multi-stage settings. The NHEFS data and the smoking cessation data are analyzed in Section \ref{DA} for illustration of the proposed method. Conclusions and discussions are provided in the last section. 
	
	\section{Q-learning for Binary Outcomes} \label{qlearn}
	
	\subsection{Notations and Assumptions}

	Without loss of generality, our attention for the model framework is restricted to a two-stage setting in Q-learning. The observed data trajectory for a patient is denoted as \{$X_{1}$, $A_{1}$, $X_{2}$, $A_{2}$, $ Y $\}, where $X_{j}$ is a covariate representing the patient's characteristics, followed by a binary treatment $A_{j} \in \{-1, 1\}$ at stage $  j $ ($ j $ = 1, 2). $Y  \in \{0, 1\}  $ denotes a binary outcome measured at the end of the second stage. History $\boldsymbol{H_{j}}$  is a collection of the covariates and treatments prior to the time of deciding the treatment $A_{j}$ at stage $ j $, $\boldsymbol{H_{1}}$ = $X_{1}$ at stage 1 and $\boldsymbol{H_{2}}$ = ($X_{1}$, $A_{1}$, $X_{2}$) at stage 2. A dynamic treatment regime is $\boldsymbol{a} $ = \{$a_{1}$, $a_{2}$\}, where $a_{j}$ = $a_{j}( \boldsymbol{h_{j}} $) is the treatment assigned at stage $ j $, and $\boldsymbol{h_{j}}$ is the observed value of $\boldsymbol{H_{j}}$. An optimal DTR is denoted as $ \boldsymbol{a^{opt}} $ = \{$a^{opt}_{1}$, $a^{opt}_{2}$\}, where $a_{j}^{opt}$ = $a_{j}^{opt}( \boldsymbol{h_{j}} $) is the optimal treatment at stage $ j $. To ensure the feasibility of Q-learning, the following assumptions are made \citep{moodie2014q}:  
	
	(A1) \textit{Stable unit treatment value}: an individual's outcome is not influenced by other individuals' treatment allocation.   
	
	(A2) \textit{No unmeasured confounders}: for any possible treatment rule, treatment $A_{j}$ received in the $ j $-th stage is independent of any future (potential) covariate or outcome conditional on the history $\boldsymbol{H_{j}}$.   
	
	(A3) \textit{Positivity}: 0 $< P(Y = 1|\boldsymbol{H_{2}}, A_{2}) <$ 1.

	\subsection{Q-learning} \label{Qlearn}
	Q-learning originates from reinforcement learning and has become one of the most popular regression-based methods to estimate an optimal DTR \citep{watkins1989learning, chakraborty2014dynamic}. The Q-learning is modeled by stage-specific Q-functions, which measure the expected future reward conditional on the history of a patient's characteristics and treatment action \citep{chakraborty2013statistical}. 
	
	When the outcome $ Y $ is binary, the Q-function at stage $ j $ can be modeled through a generalized linear model \citep{moodie2014q}
	\begin{equation} \label{qfuntrue}
		\begin{split}
			Q_{2}(\boldsymbol{H_{2}}, A_{2}; \boldsymbol{\beta_{2}}, \boldsymbol{\psi_{2}}) & = E[Y | \boldsymbol{H_{2}}, A_{2}] = \text{expit} \big(\boldsymbol{\beta_{2}^{T}H_{20}} + (\boldsymbol{\psi_{2}^{T}H_{21}})A_{2}\big),  \\
			Q_{1}(\boldsymbol{H_{1}}, A_{1}; \boldsymbol{\beta_{1}}, \boldsymbol{\psi_{1}}) & = \text{expit}\big(\boldsymbol{\beta_{1}^{T}H_{10}} + (\boldsymbol{\psi_{1}^{T}H_{11}})A_{1}\big),  
		\end{split}
	\end{equation}
	where expit($ x $) = 1/(1 + exp(-$ x $)). The term $\boldsymbol{\beta_{j}^{T}}\boldsymbol{H_{j0}}$ that does not interact with the current stage treatment $A_{j}$ is the treatment-free component, and the term $\boldsymbol{\psi_{j}^{T}}\boldsymbol{H_{j1}}$ that interacts with $A_{j}$ is the blip component, depending $ \boldsymbol{H_{j0}} $ or $ \boldsymbol{H_{j1}} $, subsets of history $\boldsymbol{H_{j}}$, respectively. The covariates in $\boldsymbol{H_{j1}}$ are called tailoring variables. 
	
	The Q-functions are usually unknown and can be estimated from the data using a backward recursive procedure \citep{chakraborty2013statistical}. In (\ref{qfuntrue}), the stage 2 parameters ($\boldsymbol{\beta_{2}}, \boldsymbol{\psi_{2}}$) can be estimated using the logistic regression. The estimation of the stage 1 parameters ($\boldsymbol{\beta_{1}}, \boldsymbol{\psi_{1}}$) relies on a pseudo-outcome denoted as $\widetilde{Y}_{1} = \underset{a_{2}}{\text{max}} \  \text{logit}Q_{2}(\boldsymbol{H_{2}},  a_{2}; \boldsymbol{\hat{\beta}_{2}}, \boldsymbol{\hat{\psi}_{2}}) $, which is the logit of the predicted probability had the patients received their second stage optimal treatment. By applying the ordinary least squares with the pseudo-outcome $\widetilde{Y}_{1}$, the first stage parameter estimates ($\boldsymbol{\hat{\beta}_{1}}, \boldsymbol{\hat{\psi}_{1}}$) are obtained.
	
	The estimation of the DTR methods generally focuses on the blip parameters, which directly decide the optimal DTR. At stage $ j $, the optimal treatment $\hat{a}_{j}^{opt}$ estimated from (\ref{qfuntrue}) can be derived by directly maximizing $Q_{j}(\boldsymbol{h_{j}}, a_{j};\boldsymbol{\hat{\beta}_{j}}, \boldsymbol{\hat{\psi}_{j}})$, $\hat{a}_{j}^{opt}$ = $\underset{a_{j}}{\arg\max} \ Q_{j}(\boldsymbol{h_{j}}, a_{j}; \boldsymbol{\hat{\beta}_{j}}, \boldsymbol{\hat{\psi}_{j}})$. As the expit function is strictly increasing, the estimated optimal treatment can also be obtained by maximizing the blip component $(\boldsymbol{\hat{\psi}_{j}^{T}}\boldsymbol{h_{j1}})a_{j}$. That is, $\hat{a}_{j}^{opt}$ = 1 if $ \boldsymbol{\hat{\psi}_{j}^{T}}\boldsymbol{h_{j1}} > 0 $, and $\hat{a}_{j}^{opt}$ = -1 otherwise.
	
	The Q-learning algorithm with a binary outcome consists of the following steps \citep{moodie2014q}:  \label{QlearnB}
	
	1. Parameterize the stage 2 Q-function 
	\begin{center}
		$Q_{2}(\boldsymbol{H_{2}}, A_{2}; \boldsymbol{\beta_{2}}, \boldsymbol{\psi_{2}}) = E[Y | \boldsymbol{H_{2}}, A_{2}] = \text{expit} \big(\boldsymbol{\beta_{2}^{T}H_{20}} + (\boldsymbol{\psi_{2}^{T}H_{21}})A_{2}\big)$. 
	\end{center}
	
	2. Apply logistic regression to obtain the stage 2 estimator ($\boldsymbol{\hat{\beta}_{2}}$, $\boldsymbol{\hat{\psi}_{2}}$).  
	
	3. Derive the stage 2 optimal treatment as $\hat{a}_{2}^{opt}$ = $\underset{a_{2}}{\arg\max} \ Q_{2}(\boldsymbol{h_{2}}, a_{2}; \boldsymbol{\hat{\beta}_{2}}, \boldsymbol{\hat{\psi}_{2}})$. 
	
	4. Construct the pseudo-outcome for stage 1 $\widetilde{Y}_{1}$ = $ \underset{a_{2}}{\text{max}} \ \text{logit}Q_{2}(\boldsymbol{H_{2}}, a_{2}; \boldsymbol{\hat{\beta}_{2}}, \boldsymbol{\hat{\psi}_{2}})$.  
	
	5. Apply ordinary least squares regression to obtain the stage 1 estimator ($\boldsymbol{\hat{\beta}_{1}}$, $\boldsymbol{\hat{\psi}_{1}}$)
	\begin{center}
		$(\boldsymbol{\hat{\beta}_{1}}, \boldsymbol{\hat{\psi}_{1}})$ = $\underset{(\boldsymbol{\beta_{1}}, \boldsymbol{\psi_{1}})}{\arg\min}$  $\frac{1}{n}$ $\sum_{i=1}^{n}$ $\big(\widetilde{Y}_{i1} - \text{logit}Q_{1}(\boldsymbol{H_{i1}}, A_{i1};\boldsymbol{\beta_{1}}, \boldsymbol{\psi_{1}})\big)^{2}$. 
	\end{center}
	
	6. Derive the stage 1 optimal treatment as $\hat{a}_{1}^{opt}$ = $\underset{a_{1}}{\arg\max} \ Q_{1}(\boldsymbol{h_{1}}, a_{1}; \boldsymbol{\hat{\beta}_{1}}, \boldsymbol{\hat{\psi}_{1}})$.

	\section{Q-learning with Misclassified Binary Outcome} \label{qlearnMis}
	
	When the binary outcome is misclassified, such as in the smoking cessation study, the Q-learning could be also affected. Let $Y^{*} $ be the surrogate outcome, a mismeasured version of $ Y $. We consider a non-differential misclassification mechanism, which could be characterized by a set of misclassification rates ($ \gamma_{10}, \gamma_{01} $) to associate $Y^{*} $ with $ Y $ such that
	\begin{equation} \label{mp2}
		\gamma_{10}  = P(Y^{*} = 1|Y = 0), \ \gamma_{01} = P(Y^{*} = 0|Y = 1),
	\end{equation}
	where the probability of $Y^{*} $ taking the value 1 depends only on the value of $ Y $. 
	
	We focus on a situation where the study of size $n$ has both internal validation data $ \mathcal{V} $ of size $n_{v}$ and main study data $\mathcal{\overline{V}}$ of size ($ n $ - $n_{v}$) available. Let subscript $ i $ represent the $ i $-th patient ($ i $ = 1, ..., $ n $). Then,
	\begin{center}
		$\{X_{i1}, A_{i1}, X_{i2}, A_{i2}, Y_{i}, Y^{*}_{i}\}$  \ \ for $i = 1, ..., n_{v}$,    \\
		$\{X_{i1}, A_{i1}, X_{i2}, A_{i2}, Y^{*}_{i}\}$ \ \ for $i = n_{v} + 1, ...,  n $,    
	\end{center}
	where the surrogate outcome $Y^{*}$ is observed for all individuals ($ i $ = 1, ..., $ n $), but the true outcome $ Y $ is only observed for individuals in the validation data ($ i $ = 1, ..., $n_{v}$).   
	
	In order for the misclassification rates and regression parameters ($ \boldsymbol{\beta_{j}} $, $\boldsymbol{\psi_{j}}$) in Q-learning to be identifiable, an additional assumption needs to be imposed: 
	
	(A4) \textit{Monotonicity condition}: $\gamma_{10}$ + $\gamma_{01} <$ 1. 
	
	The assumption (A4) ensures that ($\gamma_{10}$, $\gamma_{01}$, $\boldsymbol{\beta_{j}}$, $\boldsymbol{\psi_{j}}$) are identifiable if E[$X_{j}X_{j}^{T}$] exists and is non-singular for $ j $ = 1, 2 \citep{hausman1998misclassification}. Otherwise, if $\gamma_{10}$ + $\gamma_{01} \ge$ 1, the identifiability is not guaranteed, and $Y^{*}$ is deemed not to occur by chance \citep{hausman1998misclassification, neuhaus1999bias}.  
	
	When the outcome misclassification is ignored, and $Y^{*}$ is used to estimate the Q-function, we obtain a naive model
	\begin{equation} \label{qfunBn2}
		\begin{split}
			Q_{2}(\boldsymbol{H_{2}}, A_{2}; \boldsymbol{\beta_{2}^{n}}, \boldsymbol{\psi_{2}^{n}}) & = E[Y^{*} | \boldsymbol{H_{2}}, A_{2}] = P(Y^{*} = 1|\boldsymbol{H_{2}}, A_{2}) = \text{expit} \big(\boldsymbol{\beta_{2}^{nT}}\boldsymbol{H_{20}} + (\boldsymbol{\psi_{2}^{nT}}\boldsymbol{H_{21}})A_{2}\big). 
		\end{split} 
	\end{equation} 
	
	It has been discussed in the literature that ignoring the misclassification in the response may result in attenuated covariate effects and a change in the model structure \citep{carroll2006measurement, yi2017statistical, neuhaus1999bias}. Thus, using the naive model ($\ref{qfunBn2}$) yields a naive estimator  ($\boldsymbol{\hat{\beta}_{2}^{n}}, \boldsymbol{\hat{\psi}_{2}^{n}}$), which may be biased from ($\boldsymbol{\beta_{2}}, \boldsymbol{\psi_{2}}$). Moreover, a biased naive estimator may further affect the first stage parameter estimation and the determination of optimal DTR. Such potential issues motivate us to search for an effective approach to accommodate the outcome misclassification effect in Q-learning.

\section{Maximum Likelihood Method} \label{method}

When the outcome is subject to misclassification, the Q-learning algorithm in Section \ref{Qlearn} needs modifications to produce consistent estimates of the parameters. We propose a maximum likelihood estimation (MLE) method for Q-learning in the internal validation/main data context. The main idea of the MLE method is to derive likelihood functions for the validation data and main study data and then combine them for a total likelihood for parameter estimation in Q-learning. Given ($\boldsymbol{H_{2}} = \boldsymbol{h_{2}}$, $A_{2} = a_{2}$), we can establish a relationship of the conditional probability of the surrogate outcome with the conditional probability of the true outcome in stage 2 as
\begin{equation} \label{mle_prob}
	\begin{split}
		P(Y^{*} = 1|\boldsymbol{H_{2}} = \boldsymbol{h_{2}}, A_{2} = a_{2}) = \gamma_{10} + (1 - \gamma_{10} - \gamma_{01})P(Y = 1|\boldsymbol{H_{2}} = \boldsymbol{h_{2}}, A_{2} = a_{2}).
	\end{split}
\end{equation}

Based on (\ref{mle_prob}), we first derive the likelihood function for patients in the validation data. For any $ i $-th patient ($i = 1, ..., n_{v}$) in the validation subset, the likelihood that involves both $Y^{*}_{i}$ and $Y_{i}$ is formed as
\begin{align*}
	L_{i} & = P(Y_{i}^{*} = y^{*}_{i}, Y_{i} = y_{i}|\boldsymbol{H_{i2}} = \boldsymbol{h_{i2}}, A_{i2} = a_{i2}) \\
	& = P(Y_{i}^{*} = y^{*}_{i}| Y_{i} = y_{i}, \boldsymbol{H_{i2}} = \boldsymbol{h_{i2}}, A_{i2} = a_{i2})P(Y_{i} = y_{i} | \boldsymbol{H_{i2}} = \boldsymbol{h_{i2}}, A_{i2} = a_{i2}) \\
	& = P(Y_{i}^{*} = y^{*}_{i}| Y_{i} = y_{i})P(Y_{i} = y_{i} | \boldsymbol{H_{i2}} = \boldsymbol{h_{i2}}, A_{i2} = a_{i2}).  
\end{align*}
Then, the corresponding likelihood $L_{v}$ across $n_{v}$ patients in the validation data follows
\begin{align*}
	L_{v} & = \prod_{i=1}^{n_{v}} L_{i} = \prod_{i=1}^{n_{v}} P(Y_{i}^{*} = y^{*}_{i}| Y_{i} = y_{i})P(Y_{i} = y_{i} | \boldsymbol{H_{i2}} = \boldsymbol{h_{i2}}, A_{i2} = a_{i2}) \\
	& = \prod_{i=1}^{n_{v}} \Bigg\{\Big[P(Y_{i}^{*} = 1| Y_{i} = 1)P(Y_{i} = 1|\boldsymbol{H_{i2}} = \boldsymbol{h_{i2}}, A_{i2} = a_{i2})\Big]^{y^{*}_{i} = 1, y_{i} = 1} \times \\
	& \ \ \ \ \ \ \ \ \ \ \ \Big[P(Y_{i}^{*} = 1| Y_{i} = 0)P(Y_{i} = 0 | \boldsymbol{H_{i2}} = \boldsymbol{h_{i2}}, A_{i2} = a_{i2})\Big]^{y^{*}_{i} = 1, y_{i} = 0} \times \\
	& \ \ \ \ \ \ \ \ \ \ \ \Big[P(Y_{i}^{*} = 0| Y_{i} = 1)P(Y_{i} = 1|\boldsymbol{H_{i2}} = \boldsymbol{h_{i2}}, A_{i2} = a_{i2})\Big]^{y^{*}_{i} = 0, y_{i} = 1} \times \\
	& \ \ \ \ \ \ \ \ \ \ \ \Big[P(Y_{i}^{*} = 0| Y_{i} = 0)P(Y_{i} = 0 | \boldsymbol{H_{i2}} = \boldsymbol{h_{i2}}, A_{i2} = a_{i2})\Big]^{y^{*}_{i} = 0, y_{i} = 0} \Bigg\} \\
	& = \prod_{i=1}^{n_{v}}\Bigg\{\Big[(1 - \gamma_{01})P(Y_{i} = 1|\boldsymbol{H_{i2}} = \boldsymbol{h_{i2}}, A_{i2} = a_{i2})\Big]^{y_{i}^{*}y_{i}} \times \\
	& \ \ \ \ \ \ \ \ \ \ \  \Big[\gamma_{10}P(Y_{i} = 0|\boldsymbol{H_{i2}} = \boldsymbol{h_{i2}}, A_{i2} = a_{i2})\Big]^{y_{i}^{*}(1 - y_{i})} \times \\
	& \ \ \ \ \ \ \ \ \ \ \ \Big[\gamma_{01}P(Y_{i} = 1|\boldsymbol{H_{i2}} = \boldsymbol{h_{i2}}, A_{i2} = a_{i2})\Big]^{(1 - y_{i}^{*})y_{i}} \times \\
	& \ \ \ \ \ \ \ \ \ \ \  \Big[(1 - \gamma_{10})P(Y_{i} = 0|\boldsymbol{H_{i2}} = \boldsymbol{h_{i2}}, A_{i2} = a_{i2})\Big]^{(1 - y_{i}^{*})(1 - y_{i})} \Bigg\}  \\
	& = \prod_{i=1}^{n_{v}}\Bigg\{\Big[(1 - \gamma_{01})P(Y_{i} = 1|\boldsymbol{H_{i2}} = \boldsymbol{h_{i2}}, A_{i2} = a_{i2})\Big]^{y_{i}^{*}y_{i}} \times \\
	& \ \ \ \ \ \ \ \ \ \ \ \Big[\gamma_{10}\Big(1 - P(Y_{i} = 1|\boldsymbol{H_{i2}} = \boldsymbol{h_{i2}}, A_{i2} = a_{i2})\Big)\Big]^{y_{i}^{*}(1 - y_{i})} \times \\
	& \ \ \ \ \ \ \ \ \ \ \ \Big[\gamma_{01}P(Y_{i} = 1|\boldsymbol{H_{i2}} = \boldsymbol{h_{i2}}, A_{i2} = a_{i2})\Big]^{(1 - y_{i}^{*})y_{i}} \times \\
	& \ \ \ \ \ \ \ \ \ \ \ \Big[(1 - \gamma_{10})\Big(1 - P(Y_{i} = 1|\boldsymbol{H_{i2}} = \boldsymbol{h_{i2}}, A_{i2} = a_{i2})\Big)\Big]^{(1 - y_{i}^{*})(1 - y_{i})} \Bigg\}. 
\end{align*}

For any $ i $-th patient in the main study data where only $Y^{*}_{i}$ is observed ($i$ = $n_{v}$ + 1, ..., $n$), the likelihood $L_{i}$ is given by
\begin{align*}
	L_{i} & = P(Y_{i}^{*} = y^{*}_{i}|\boldsymbol{H_{i2}} = \boldsymbol{h_{i2}}, A_{i2} = a_{i2}) \\
	& = P(Y_{i}^{*} = y^{*}_{i}, Y_{i} = 1|\boldsymbol{H_{i2}} = \boldsymbol{h_{i2}}, A_{i2} = a_{i2}) + P(Y_{i}^{*} = y^{*}_{i}, Y_{i} = 0|\boldsymbol{H_{i2}} = \boldsymbol{h_{i2}}, A_{i2} = a_{i2}) \\
	& = P(Y_{i}^{*} = y^{*}_{i}| Y_{i} = 1, \boldsymbol{H_{i2}} = \boldsymbol{h_{i2}}, A_{i2} = a_{i2})P(Y_{i} = 1|\boldsymbol{H_{i2}} = \boldsymbol{h_{i2}}, A_{i2} = a_{i2}) +  \\
	& \ \ \ \ P(Y_{i}^{*} = y^{*}_{i}| Y_{i} = 0, \boldsymbol{H_{i2}} = \boldsymbol{h_{i2}}, A_{i2} = a_{i2})P(Y_{i} = 0 | \boldsymbol{H_{i2}} = \boldsymbol{h_{i2}}, A_{i2} = a_{i2})  \\
	& = P(Y_{i}^{*} = y^{*}_{i}| Y_{i} = 1)P(Y_{i} = 1|\boldsymbol{H_{i2}} = \boldsymbol{h_{i2}}, A_{i2} = a_{i2}) + \\
	& \ \ \ \  P(Y_{i}^{*} = y^{*}_{i}| Y_{i} = 0)P(Y_{i} = 0 | \boldsymbol{H_{i2}} = \boldsymbol{h_{i2}}, A_{i2} = a_{i2}).  
\end{align*}
Then, the likelihood $L_{\overline{v}}$ is the product of the likelihoods across ($n$ - $n_{v}$) patients from the main study data
\begin{align*}
	L_{\overline{v}} & = \prod_{i=n_{v}+1}^{n} L_{i} \\
	& = \prod_{i=n_{v}+1}^{n} \bigg\{P(Y_{i}^{*} = y^{*}_{i}| Y_{i} = 1)P(Y_{i} = 1|\boldsymbol{H_{i2}} = \boldsymbol{h_{i2}}, A_{i2} = a_{i2}) + \\
	& \ \ \ \ \ \ \ \ \ \ \ \ \ \ P(Y_{i}^{*} = y^{*}_{i}| Y_{i} = 0)P(Y_{i} = 0 | \boldsymbol{H_{i2}} = \boldsymbol{h_{i2}}, A_{i2} = a_{i2}) \bigg\} \\
	& = \prod_{i=n_{v}+1}^{n} \bigg\{P(Y_{i}^{*} = 1| Y_{i} = 1)P(Y_{i} = 1|\boldsymbol{H_{i2}} = \boldsymbol{h_{i2}}, A_{i2} = a_{i2}) + \\
	& \ \ \ \ \ \ \ \ \ \ \ \ \ \ P(Y_{i}^{*} = 1| Y_{i} = 0)P(Y_{i} = 0 | \boldsymbol{H_{i2}} = \boldsymbol{h_{i2}}, A_{i2} = a_{i2}) \bigg\}^{y_{i}^{*}}  \times \\
	& \ \ \ \ \ \ \ \ \ \ \ \ \bigg\{P(Y_{i}^{*} = 0| Y_{i} = 1)P(Y_{i} = 1|\boldsymbol{H_{i2}} = \boldsymbol{h_{i2}}, A_{i2} = a_{i2}) + \\
	& \ \ \ \ \ \ \ \ \ \ \ \ \ \ P(Y_{i}^{*} = 0| Y_{i} = 0)P(Y_{i} = 0 | \boldsymbol{H_{i2}} = \boldsymbol{h_{i2}}, A_{i2} = a_{i2}) \bigg\}^{1-y_{i}^{*}}  \\
	& = \prod_{i=n_{v}+1}^{n} \bigg\{(1 - \gamma_{01})P(Y_{i} = 1|\boldsymbol{H_{i2}} = \boldsymbol{h_{i2}}, A_{i2} = a_{i2}) + \gamma_{10}P(Y_{i} = 0|\boldsymbol{H_{i2}} = \boldsymbol{h_{i2}}, A_{i2} = a_{i2})\bigg\}^{y_{i}^{*}} \times \\
	& \ \ \ \ \ \ \ \ \ \ \ \ \bigg\{\gamma_{01}P(Y_{i} = 1|\boldsymbol{H_{i2}} = \boldsymbol{h_{i2}}, A_{i2} = a_{i2}) + (1 - \gamma_{10})P(Y_{i} = 0|\boldsymbol{H_{i2}} = \boldsymbol{h_{i2}}, A_{i2} = a_{i2})\bigg\}^{1 - y_{i}^{*}} \\
	& = \prod_{i=n_{v}+1}^{n} \bigg\{(1 - \gamma_{01})P(Y_{i} = 1|\boldsymbol{H_{i2}} = \boldsymbol{h_{i2}}, A_{i2} = a_{i2}) + \gamma_{10}\Big[1 - P(Y_{i} = 1|\boldsymbol{H_{i2}} = \boldsymbol{h_{i2}}, A_{i2} = a_{i2})\Big]\bigg\}^{y_{i}^{*}} \times \\
	& \ \ \ \ \ \ \ \ \ \ \ \ \bigg\{\gamma_{01}P(Y_{i} = 1|\boldsymbol{H_{i2}} = \boldsymbol{h_{i2}}, A_{i2} = a_{i2}) + (1 - \gamma_{10})\Big[1 - P(Y_{i} = 1|\boldsymbol{H_{i2}} = \boldsymbol{h_{i2}}, A_{i2} = a_{i2})\Big]\bigg\}^{1 - y_{i}^{*}} \\
	& = \prod_{i=n_{v}+1}^{n} \bigg\{\gamma_{10} + (1 - \gamma_{10} - \gamma_{01})P(Y_{i} = 1|\boldsymbol{H_{i2}} = \boldsymbol{h_{i2}}, A_{i2} = a_{i2}) \bigg\}^{y_{i}^{*}} \times  \\
	& \ \ \ \ \ \ \ \ \ \ \ \ \bigg\{(1 - \gamma_{10}) - (1 - \gamma_{10} - \gamma_{01})P(Y_{i} = 1|\boldsymbol{H_{i2}} = \boldsymbol{h_{i2}}, A_{i2} = a_{i2}) \bigg\}^{1 - y_{i}^{*}}. 
\end{align*}

Thus, the total likelihood function $ L $ for all patients from both the validation study and the main study is given by 
\begin{equation} \label{likeliF}
	L = L_{v} \times L_{\overline{v}}, 
\end{equation}
and a total log-likelihood function that is to be maximized is expressed as
\begin{equation} \label{loglikeliF}
	\begin{split}
		logL & = \sum_{i=1}^{n_{v}} \Bigg\{y_{i}^{*}y_{i}log\Big[(1 - \gamma_{01})P(Y_{i} = 1|\boldsymbol{H_{i2}} = \boldsymbol{h_{i2}}, A_{i2} = a_{i2})\Big] + \\
		& \ \ \ \ \ \ \ \ \ \ \ \ \ \  y_{i}^{*}(1 - y_{i})log\Big[\gamma_{10}\big(1 - P(Y_{i} = 1|\boldsymbol{H_{i2}} = \boldsymbol{h_{i2}}, A_{i2} = a_{i2})\big)\Big] + \\
		& \ \ \ \ \ \ \ \ \ \ \ \ \ \  (1 - y_{i}^{*})y_{i}log\Big[\gamma_{01}P(Y_{i} = 1|\boldsymbol{H_{i2}} = \boldsymbol{h_{i2}}, A_{i2} = a_{i2})\Big] + \\
		& \ \ \ \ \ \ \ \ \ \ \ \ \ \  (1 - y_{i}^{*})(1 - y_{i})log\Big[(1 - \gamma_{10})\big(1 - P(Y_{i} = 1|\boldsymbol{H_{i2}} = \boldsymbol{h_{i2}}, A_{i2} = a_{i2})\big)\Big] \Bigg\} + \\
		& \ \ \ \ \sum_{i=n_{v}+1}^{n} \Bigg\{y_{i}^{*}log\Big[\gamma_{10} + (1 - \gamma_{10} - \gamma_{01})P(Y_{i} = 1|\boldsymbol{H_{i2}} = \boldsymbol{h_{i2}}, A_{i2} = a_{i2}) \Big] + \\
		& \ \ \ \ \ \ \ \ \ \ \ \ \ \ \ \  (1 - y_{i}^{*})log\Big[(1 - \gamma_{10}) - (1 - \gamma_{10} - \gamma_{01})P(Y_{i} = 1|\boldsymbol{H_{i2}} = \boldsymbol{h_{i2}}, A_{i2} = a_{i2}) \Big]\Bigg\}.  	\end{split}
\end{equation}

Maximizing $logL$ (\ref{loglikeliF}) with respect to $\boldsymbol{\theta}$ = ($ \boldsymbol{\beta_{2}} $, $ \boldsymbol{\psi_{2}} $, $ \gamma_{10} $, $ \gamma_{01} $) results in a MLE estimator $\boldsymbol{\hat{\theta}^{mle}}$ for $\boldsymbol{\theta}$. Once the stage 2 estimator ($ \boldsymbol{\hat{\beta}^{mle}_{2}} $, $ \boldsymbol{\hat{\psi}^{mle}_{2}} $) is obtained, the pseudo-outcome is reconstructed using the ($ \boldsymbol{\hat{\beta}^{mle}_{2}} $, $ \boldsymbol{\hat{\psi}^{mle}_{2}} $), and ordinary least squares is applied to find the stage 1 estimator ($ \boldsymbol{\hat{\beta}^{mle}_{1}} $, $ \boldsymbol{\hat{\psi}^{mle}_{1}} $). 

%\begin{theorem}
	Under suitable conditions (C1) - (C5) in the Appendix, the MLE estimator $\boldsymbol{\hat{\theta}^{mle}}$ for stage 2 is a consistent estimator of $\boldsymbol{\theta}$. That is,
	\begin{align*}
		\boldsymbol{\hat{\theta}^{mle}} \overset{p}{\to} \boldsymbol{\theta} \ \ \ as \ n \rightarrow \infty.
	\end{align*}
%\end{theorem}

The consistent estimator ($ \boldsymbol{\hat{\beta}_{2}^{mle}} $, $ \boldsymbol{\hat{\psi}_{2}^{mle}} $) in stage 2 ensures the pseudo-outcome estimation to be consistent in stage 1, which further provides consistent estimator ($ \boldsymbol{\hat{\beta}_{1}^{mle}} $, $ \boldsymbol{\hat{\psi}_{1}^{mle}} $) in stage 1 using the ordinary least squares. Thus, the MLE method yields consistent estimates of blip parameter $ \boldsymbol{\psi} = (\boldsymbol{\psi_{2}}, \boldsymbol{\psi_{1}})$ in Q-learning. The detailed proofs are provided in the Appendix.   

The following two-stage Q-learning algorithm provides the modified estimation procedures:   \label{Qlearn.mle}  

1. Parameterize the stage 2 Q-function
\begin{center}
	$Q_{2}(\boldsymbol{H_{2}}, A_{2}; \boldsymbol{\beta_{2}}, \boldsymbol{\psi_{2}}) = \text{expit} \big(\boldsymbol{\beta_{2}^{T}H_{20}} + (\boldsymbol{\psi_{2}^{T}H_{21}})A_{2}\big)$.  
\end{center}

2. Apply maximum likelihood estimation method to obtain the stage 2 estimator ($\boldsymbol{\hat{\beta}_{2}^{mle}}$, $\boldsymbol{\hat{\psi}_{2}^{mle}}$) by maximizing the log-likelihood function (\ref{loglikeliF}).  

3. Derive the stage 2 optimal treatment as $\hat{a}_{2}^{opt}$ = $\underset{a_{2}}{\arg\max} \ Q_{2}(\boldsymbol{h_{2}}, a_{2}; \boldsymbol{\hat{\beta}_{2}^{mle}}, \boldsymbol{\hat{\psi}_{2}^{mle}})$.

4. Construct the pseudo-outcome at stage 1  $\widetilde{Y}_{1}$ = $\underset{a_{2}}{\text{max}} \ \text{logit}Q_{2}(\boldsymbol{H_{2}}, a_{2}; \boldsymbol{\hat{\beta}_{2}^{mle}}, \boldsymbol{\hat{\psi}_{2}^{mle}})$. 

5. Apply ordinary least squares regression to obtain the stage 1 estimator ($\boldsymbol{\hat{\beta}_{1}^{mle}}$, $\boldsymbol{\hat{\psi}_{1}^{mle}}$)
\begin{center}
	$(\boldsymbol{\hat{\beta}^{mle}_{1}}, \boldsymbol{\hat{\psi}^{mle}_{1}}) = \underset{(\boldsymbol{\beta_{1}},  \boldsymbol{\psi_{1}})}{\arg\min}  \frac{1}{n} \sum_{i=1}^{n} \big(\widetilde{Y}_{i1} - \text{logit}Q_{1}(\boldsymbol{H_{i1}}, A_{i1};\boldsymbol{\beta_{1}}, \boldsymbol{\psi_{1}})\big)^{2}$.
\end{center}

6. Derive the stage 1 optimal treatment as $\hat{a}_{1}^{opt}$ = $\underset{a_{1}}{\arg\max} \ Q_{1}(\boldsymbol{h_{1}}, a_{1}; \boldsymbol{\hat{\beta}_{1}^{mle}}, \boldsymbol{\hat{\psi}_{1}^{mle}})$. 

This modified Q-learning algorithm distinguishes itself from the original Q-learning algorithm in Step 2, which replaces the application of logistic regression with the maximum likelihood estimation method.

\section{SIMULATION STUDY} \label{sim}

Simulation studies are conducted to assess the performance of the proposed Q-learning method in different scenarios by assessing parameter estimation, prediction accuracy of optimal DTR and predicted optimal outcome.

\subsection{One-Stage Estimation} \label{one_est} 

We begin with the one-stage estimation in Q-learning. Let $ X $ be a continuous covariate and $ Z $ be a binary covariate, where $X \sim N(1, 1)$ and $Z \in \{-1, 1\}$ is generated with $ P $($ Z  $ = 1) = 0.5. The treatment $A \in \{-1, 1\}$ is drawn from a Bernoulli distribution with probability$  P$($A $ = 1) = expit(1 - $ X $). The true outcome $ Y $ is drawn from a Bernoulli distribution with probability expit\big(1 + $\beta_{z}Z$ + $\beta_{x}X$ + ($\psi_{10}$ + $\psi_{11}X$)$A$\big), where ($\boldsymbol{\beta}$, $\boldsymbol{\psi}$) = ($\beta_{z}, \beta_{x}, \psi_{10}, \psi_{11}$) = (0.5, -1, 0.5, -0.5). Misclassified outcome $Y^{*}$ is simulated from a Bernoulli distribution based on the specified misclassification rates ($\gamma_{10}, \gamma_{01}$). 

The generated dataset is randomly divided into validation data and main study data with a validation ratio $\rho $, where the validation data contain 100$\times \rho \%$ of the observations. We consider three estimators to evaluate the performance of the proposed method: (1) validation estimator ($\boldsymbol{\hat{\beta}^{v}}$, $\boldsymbol{\hat{\psi}^{v}}$) obtained using the validation data only, (2) naive estimator ($\boldsymbol{\hat{\beta}^{n}}$, $\boldsymbol{\hat{\psi}^{n}}$) obtained using the surrogate outcome $Y^{*}$, (3) MLE estimator ($\boldsymbol{\hat{\beta}^{mle}}$, $\boldsymbol{\hat{\psi}^{mle}}$) obtained from the modified algorithm in Section \ref{Qlearn.mle}. 

We compare results under two sample sizes of $ n $ = 500 and $ n $ = 2000. The validation ratio $\rho $ is specified as 0.3 and 0.5. The misclassification rates of ($\gamma_{10}, \gamma_{01}$) is set to be (0.1, 0.1), (0.2, 0.2) and (0.3, 0.3). Simulations are repeated 500 times for all the combinations of $ \rho$ and ($\gamma_{10}, \gamma_{01}$). The average bias, empirical standard error (SE), and root mean square error (RMSE) of $\boldsymbol{\hat{\psi}}$ are reported. The percentile bootstrap confidence intervals are also calculated with 200 bootstrap samples to derive the coverage rate (CR\%) of 95\% confidence intervals. The numerical results are summarized in Table $\ref{tab:one}$. 

Table $\ref{tab:one}$ shows that ignoring the outcome misclassification, the naive estimator $\boldsymbol{\hat{\psi}^{n}}$ produces biased results. The biases become servere as the misclassification rate increases. On the contrary, the proposed estimator yields small biases and the coverage rates are close to the nominal level  95\%. Moreover, the proposed method is numerically stable and robust against different settings of $\rho$ and ($\gamma_{10}, \gamma_{01}$). The sample size also plays an important role in the performance of methods. As $\rho$ or $ n $ increases, the biases and variability of the estimators are reduced.

\subsection{Two-Stage Estimation} \label{two_est} 

In this section, we extend the study to evaluate the performance of the proposed method with two decision points. For simplicity, we follow the same simulation design as in \cite{moodie2014q}, where the confounding variables are present. 

A dataset with 2000 patients forms data trajectory $(X_{1}, Z_{1}, A_{1}, X_{2}, Z_{2}, A_{2}, Y)$. For $ j $ = 1, 2, $X_{j}$ is a continuous confounding covariate at stage $ j $, where $X_{1} \sim N(0, 1)$ and $X_{2} \sim N(\eta_{0} + \eta_{1}X_{1}, 1) $  with $\eta_{0} = -0.5, \eta_{1} = 0.5$. The treatment $A_{j} \in \{-1, 1\}$ is generated with probability $ P $($A_{j}$ = 1) = expit($\zeta_{0}$ + $\zeta_{1}X_{j}$) for $\zeta_{0} = -0.8$ and $\zeta_{1} = 1.25$. Two binary covariates $Z_{j} \in \{-1, 1\}$ are generated as $ P $($Z_{1}$ = 1) = 0.5 and $ P $($Z_{2} = 1 | Z_{1}, A_{1}$) = expit$\big(\delta_{1}Z_{1} + \delta_{2}A_{1}\big)$. Given the data trajectory, the history at each stage is $\boldsymbol{H_{1}} = (X_{1}, Z_{1})$ and $\boldsymbol{H_{2}} = (X_{1}, Z_{1}, A_{1}, X_{2}, Z_{2})$. The outcome model is given by
\begin{center}
	$ P(Y = 1|\boldsymbol{H_{2}}, A_{2}; \boldsymbol{\beta}, \boldsymbol{\psi})$ = expit($\beta_{0} + \beta_{1}X_{1} + \beta_{2}Z_{1} + \beta_{3}A_{1} + \beta_{4}Z_{1}A_{1} + \beta_{5}X_{2} + \psi_{20}A_{2} + \psi_{21}Z_{2}A_{2} + \psi_{22}A_{1}A_{2}) $
\end{center}

We consider a complete regular scenario and set ($\boldsymbol{\beta}$, $\boldsymbol{\psi}$) = ($\beta_{0}$, $\beta_{1}$, $\beta_{2}$, $\beta_{3}$, $\beta_{4}$, $\beta_{5}$, $\psi_{20}$, $\psi_{21}$, $\psi_{22}$) = (0, 1, 0, -0.5, 0, 1, 0.25, 0.5, 0.5) and ($\delta_{1}$, $\delta_{2}$) = (0.1, 0.1). The stage 2 blip parameter is $\boldsymbol{\psi_{2}}$ = ($\psi_{20}$, $\psi_{21}$, $\psi_{22}$), and the stage 1 blip parameter $\boldsymbol{\psi_{1}}$ = ($\psi_{10}$, $\psi_{11}$) is quantified as $\psi_{10}$ = -0.3688 and $\psi_{11}$ = 0.0187 based on the data-generating parameters ($\boldsymbol{\beta}$, $\boldsymbol{\psi}$) in this setting \citep{moodie2014q}. The observed surrogate $Y^{*}$ is generated from a Bernoulli distribution based on the misclassification model (\ref{mp2}) after the true outcome is obtained, where the misclassification rates ($\gamma_{10}, \gamma_{01}$) are set to be (0.1, 0.1), (0.2, 0.2) and (0.3, 0.3). Once the dataset is generated, the validation data is randomly separated with a ratio $\rho \in$ \{0.3, 0.5\}. The three estimators described in section (\ref{one_est}) are involved to estimate the parameters of interest. 500 simulations are run for the combinations of $\rho$ and ($\gamma_{10}, \gamma_{01}$). Numerical results for the bias, empirical standard error (SE), root mean square error (RMSE) and 95\% coverage rate (CR\%) of $\boldsymbol{\hat{\psi}}$ = ($\boldsymbol{\hat{\psi}_{2}}$, $\boldsymbol{\hat{\psi}_{1}}$) are reported in Table \ref{tab:two}. 

Similar to the one-stage setting, Table \ref{tab:two} shows that the naive estimator $\boldsymbol{\hat{\psi}^{n}}$ leads to broadly biased results. However, the proposed estimator $\boldsymbol{\hat{\psi}^{mle}}$ outperforms the naive estimator with small biases in all the scenarios, and the coverage rates of $\boldsymbol{\hat{\psi}^{mle}}$ are close to 95\%. Moreover, the results shows that for the set of first stage estimators, $\hat{\psi}_{10}$ is generally more vulnerable to bias compared with $\hat{\psi}_{11}$, which agrees with the findings in the literature \citep{moodie2014q, chakraborty2010inference, song2015penalized}. 

\subsection{Prediction}  \label{pred1} 

We also explore the misclassification effect in a Q-learning model from a prediction perspective. We are particularly interested in assessing the prediction accuracy of optimal DTR, the predicted error rates, sensitivity, and specificity of the outcome under the optimal DTR.   

The simulation design follows (\ref{two_est}), but the simulated data consists of training data with 2000 patients and test data with 5000 patients. The training data are randomly split into validation data and main study data with $\rho \in$ \{0.3, 0.5\}, by which the misclassification rates and the regression parameters are estimated. We evaluate the performance of the proposed correction method in a predictive setting under the previous three estimators ($\boldsymbol{\hat{\beta}^{v}}$, $\boldsymbol{\hat{\psi}^{v}}$), ($\boldsymbol{\hat{\beta}^{n}}$, $\boldsymbol{\hat{\psi}^{n}}$), ($\boldsymbol{\hat{\beta}^{mle}}$, $\boldsymbol{\hat{\psi}^{mle}}$). The test data are used to compute the prediction accuracy of optimal DTR, which is measured by the proportion of patients whose optimal treatments are correctly predicted at stage 2 and/or stage 1. Then, based on the estimated optimal DTR, we calculate the (1) predicted error rates of the outcome, which is measured by the proportion of patients whose outcomes are incorrectly predicted under the estimated optimal DTR, (2) predicted sensitivity of the outcome, which is measured by the proportion of patients whose positive outcomes ($ Y $ = 1) are correctly predicted under the estimated optimal DTR, (3) predicted specificity of the outcome, which is measured by the proportion of patients whose negative outcomes ($ Y $ = 0) are correctly predicted under the estimated optimal DTR. For the training data, the validation ratio $\rho$ is specified as 0.3 and 0.5, and the misclassification rates ($\gamma_{10}, \gamma_{01}$) are set to be (0.1, 0.1), (0.2, 0.2) and (0.3, 0.3). Simulations are repeated 500 times. Results are summarized in Table \ref{tab:prediction}.

Table $\ref{tab:prediction}$A shows that the prediction accuracy of optimal DTR is adversely affected by the misclassification. The naive estimator leads to a pronounced degeneration in the accuracy of predicted optimal DTR, and its performance is worsened as the misclassification rate increases. In comparison, the proposed method considerably improves the prediction accuracy, especially when the optimal treatments in both stages are evaluated. The proposed method is also robust against the magnitudes of $\rho$ and ($\gamma_{10}, \gamma_{01}$). It substantially restores the precision to a level that is even superior to the validation estimator, suggesting a favorable alternative choice to derive the sequential optimal treatment rules.  

Table $\ref{tab:prediction}$B shows that the naive method leads to the worst results in terms of the predicted error rates, sensitivity, and specificity of the outcome in most scenarios. Moreover, compared with sensitivity, specificity is generally more sensitive to the outcome misclassification as more positive outcome values are predicted. In contrast, the proposed MLE method produces the best performance with the lowest error rates and highest sensitivity and specificity results in all the scenarios. As $\rho$ increases, the predicted error rates of the proposed method decrease, and the predicted sensitivity and specificity of the proposed method increase.

\section{Data Analysis} \label{DA}
\subsection{NHEFS Data} \label{NHEFS} 

The NHEFS study was conducted by the National Center for Health Statistics and the National Institute on Aging in collaboration with other agencies of the Public Health Service. A detailed description of the NHEFS is available at \url{https://wwwn.cdc.gov/nchs/nhanes/nhefs/}. The NHEFS study aimed to investigate the relationships between clinical, nutritional, and behavioral factors assessed in the first National Health and Nutrition Examination Survey NHANES I and subsequent morbidity, mortality, and operational factors with hospital utilization. We are interested in estimating an optimal treatment decision rule using the cohort NHEFS dataset in \cite{hernan2010causal}. The dataset consists of 1566 cigarette smokers aged 25-74 years, with a number of baseline variables collected from 1971 to 1975. They were followed up through personal interviews in 1982 and reported quitting smoking status, which is the outcome of interest in the analysis. We consider a binary indicator for regular exercise as the treatment variable, with $ A $ = 1 indicating those who had little or no exercise and $ A $ = -1 otherwise. The baseline variables to be included are age, gender, race, body mass index, systolic blood pressure (SBP), physical activity status, cholesterol, weight, diabetes, the number of years of smoking, and the number of cigarettes smoked each day (SmokeIntensity). Since the measured SBP is right-skewed in the dataset, we take the logarithmic transformation of SBP to be $ log $(SBP-50)\citep{carroll2006measurement}. Diabetes and SmokeIntensity are shown to be significantly associated with the treatment variable from the treatment model. We regard these two variables as the tailoring variables to derive the optimal treatment decision rule. All the continuous variables are standardized in the analysis.  

As described, the smoking status is reported by the patients and thus subject to misclassification. In the dataset, there is no information available to know about the true smoking status, making it difficult to infer the degree of misclassification rates. Therefore, we only include the main study data and specify a series of values for the misclassification rates and conduct sensitivity analyses to evaluate how the misclassification rate affects the estimated optimal treatment decision rule. It is discussed that the smokers who have really quit smoking are unlikely to report they are still smoking, while those who have not quit smoking are likely to misreport their smoking cessation status \citep{magder1997logistic}. In the literature, an estimate for the misclassification rate was reported as $\gamma_{10} = 7.5\%$ in a smoking cessation study \citep{lee2013effectiveness}. Thus, we consider $\gamma_{01}$ = 0 and $\gamma_{10} \in (5\%, 7.5\%, 10\%, 12.5\%)$ in our analysis. Table \ref{tab:NHEFS}A summarizes the associated results, including the estimates, bootstrap standard error (SE), and 95\% confidence intervals (CI) for the blip parameters obtained from the naive method and the proposed method. 

From Table \ref{tab:NHEFS}A, we see the estimated optimal treatment decision rule based on the naive method is $\hat{a}^{opt}$ = 1 if -0.148 + 0.130$\times$Diabetes + 0.075$\times$SmokeIntensity $>$ 0, and $\hat{a}^{opt}$ = -1 otherwise. In general, the proposed method produces slightly larger estimates than the naive method, leading to different optimal treatment decision rules. As $\gamma_{10}$ increases, the blip parameter estimates and estimated SEs obtained from the proposed method become bigger. Moreover, the diabetes variable is shown to have a significant treatment effect  in the naive method, but the MLE method displays different statistical significance for diabetes in all the scenarios. Therefore, it reveals that the misclassification effect is not negligible in an error-prone setting, which can alter the decision results, including the statistical significance, when the misclassification is taken into account in the analysis.

\subsection{Smoking Cessation Data} \label{SCD}

In the second example, we explore the misclassification effect by analyzing the smoking cessation data, which were collected at St. Joseph's Hospital \citep{lee2013effectiveness}. The smoking cessation study is a randomized controlled trial and aims to examine the effectiveness of a perioperative smoking cessation intervention with one decision point involved. We are interested in using the smoking cessation data to estimate an optimal treatment decision rule. In this trial, 168 patients were recruited and randomly assigned with the probability 0.5 to one of the two treatment groups, the intervention group ($ A $ = 1) or the control group ($ A $ = -1). The patients were followed up at the time of the 30-day postoperative phone call and self-reported smoking cessation status were obtained, which is the outcome of interest with $ Y $ = 1 indicating smoking cessation.   

In the study, the smoking cessation status reported by the smokers was examined with the exhaled carbon monoxide (CO) levels (ppm), where an exhaled CO of $\le$ 10 ppm confirmed smoking quitting \citep{lee2013effectiveness}. It has been found that out of 146 patients with exhaled CO greater than 10ppm, 11 patients misreported their smoking cessation status. With a non-differential misclassification mechanism assumed, the misclassification rate can be estimated as $\gamma_{10}$ = 11/146 = 7.5\%. For those who have already quit smoking, it has been shown that they were highly likely to report that they have stopped smoking \citep{magder1997logistic}. Then, we assume that  $\gamma_{01}$ = 0. It should be noted that these ($\gamma_{10} $, $ \gamma_{01} $) are the estimates of misclassification rates while the true misclassification rates are still unknown. Thus, we take a series of values for $\gamma_{10} \in (2.5\%, 5\%, 7.5\%, 8.5\%)$ and conduct sensitivity analyses to evaluate how the misclassification rate affects the optimal treatment decision rule estimation. The baseline variables in the analysis include age, gender, body mass index, diabetes status, hypertension, cigarettes smoked per day, and the number of years of smoking. The hypertension variable was found statistically significant with respect to the treatment \citep{shu2019causal}. We consider hypertension (HTN) and the number of years of smoking (YrsSmoke) as the tailoring variables to derive the optimal treatment decision rule. All the continuous variables are standardized in the analysis. Table $\ref{tab:smoking}$B summarizes the inference results obtained from the naive method and the proposed method.

The analysis results suggest that the misclassification effect is conspicuous. The naive method leads to an optimal decision rule, which is determined by the values of (0.628 - 0.244$\times$HTN - 0.162$\times$YrsSmoke). In comparison, the proposed MLE method yields notably larger parameter estimates and estimated standard errors than the naive method. Moreover, we observe again that the significant of the treatment variable is changed when the misclassification is taken into account. As $\gamma_{10}$ increases, the MLE estimator is sensitive to the change in the misclassification rate. One possible reason might be the limited size of the dataset. However, these results still reveal a non-negligible impact of outcome misclassification on the optimal treatment decision rule estimation for smoking cessation.

\section{DISCUSSION} \label{conclude}

In this paper, we explore Q-learning with misclassified binary outcomes in the context of internal validation/main data design. We show that when the outcome misclassification is ignored, the parameter estimation in Q-learning is severely biased, and the optimal decision rule may be affected. We propose a correction method based on the relationship between two conditional probabilities of the true outcome and surrogate outcome. The likelihoods for both the validation and main study data are derived and combined to generate a total likelihood, which is used for parameter estimation in Q-learning. The proposed method is proved to yield consistent estimates of blip parameters in Q-learning under suitable conditions.     

Simulation studies demonstrate the biasedness in the estimation due to the outcome misclassification in Q-learning when the outcome misclassification is ignored. By making use of the observed surrogate outcome and validation data, the proposed method effectively corrects the bias and confirms the theoretical property with satisfactory finite sample performance in both single-stage and multi-stage settings. It is also robust against the various magnitude of misclassification rate and sample size. In the predictive setting, we observe a deterioration in the accuracy of correctly predicting the optimal treatments across stages if the misclassification is ignored in Q-learning. When the outcome is evaluated by the naive method, the predicted error rates are conspicuously higher and predicted sensitivity and specificity are noticeably lower. As a comparison, the proposed method rectifies the situation by enhancing the prediction accuracy of optimal DTR, predicted sensitivity and specificity of the outcome, and largely reducing the predicted error rates, whose performance is also superior to using the validation data only.

Sensitivity analyses are performed for the NHEFS data and the smoking cessation data to compare the optimal treatment decision rules estimated from the naive method and the proposed method. By incorporating the misclassification in the analysis, the estimated optimal treatment rules are shown to be different, and the statistical significance of the variables is also altered. The data analysis reveals a non-negligible impact of outcome misclassification in terms of optimal treatment decision rules estimation.

The misclassification problem in dynamic treatment regimes is a new and challenging topic. In this study, we assume that internal validation data are available in Q-learning, in which both true and surrogate outcomes are observed. It is of interest to consider the misclassification problem in Q-learning with replicate data. Instead of observing the true outcome in a small subset of data, only the replicates of the outcome are observed. In such circumstances, it is necessary to explore other approaches to correct the outcome misclassification in Q-learning.

\bibliographystyle{apalike}
\bibliography{QBinaryMis_arXiv}

\begin{appendices} 
	\section{Proof of Consistency}\label{appendix}
	The proof of consistency in this section is based on a one-stage setting, and it can be intuitively extended to multiple stages. Let $\boldsymbol{\theta}$ = ($\boldsymbol{\beta}, \boldsymbol{\psi}, \gamma_{10}, \gamma_{01}$) and $\boldsymbol{\hat{\theta}^{mle}}$ be the MLE estimator. The conditions for the property of consistency in Q-learning include: \\
	
	(C1) Let $\Omega$ be the parameter space with finite dimension for $\boldsymbol{\theta}$. $\Omega$ is closed and compact. The true parameter value of $\boldsymbol{\theta}$ is interior to $\Omega$.  
	
	(C2) The probability distributions with any two different values of $\boldsymbol{\theta}$ are distinct.  
	
	(C3) For an open subset $\omega$ of $\Omega$ that contains the true parameter value of $\boldsymbol{\theta}$, the first three derivatives of the log-likelihood \textit{l}($\boldsymbol{\theta}$) exist for $\boldsymbol{\theta} \in \omega$ almost surely. There exists a function M such that the n$^{-1}$ times the absolute value of the the third derivative is bounded by M for $\boldsymbol{\theta} \in \omega$ and E[M] $< \infty$.  
	
	(C4) The information matrix \textit{I}($\boldsymbol{\theta}$) is finite and positive definite for $\boldsymbol{\theta} \in \omega$.  
	
	(C5) Assumptions (A1) - (A4) in Sections \ref{qlearn} and \ref{qlearnMis} hold.   \\

	The conditions contain the regularity conditions (C1) - (C4) \citep{cox1979theoretical} and the assumptions that are necessary for Q-learning. The condition (C5) guarantees the identifiability of the parameter $\boldsymbol{\theta}$ in Q-learning to estimate a dynamic treatment regime. According to \cite{pepe1992inference}, under the conditions (C1) - (C5), the MLE estimator $\boldsymbol{\hat{\theta}^{mle}}$ that solves the equation $\frac{\partial}{\partial \boldsymbol{\theta}} log L(\boldsymbol{\theta}) = 0$ satisifies 
	\begin{center}
		$ 	\boldsymbol{\hat{\theta}^{mle}} \overset{p}{\to} \boldsymbol{\theta} \ \ \ as \ n \rightarrow \infty$,
	\end{center}
	where $ L(\boldsymbol{\theta}) $ is the likelihood stated in (\ref{likeliF}). Thus, $\boldsymbol{\hat{\psi}^{mle}}$ is a consistent estimator of blip parameter $ \boldsymbol{\psi} $. 
	
	\section{Tables}
	
	\begin{table} 
		\caption{One-stage estimates of blip parameters $(\psi_{10}, \psi_{11})$}
		\label{tab:one}
		\centering
		\begin{tabular}{cccrrccrrrccr}
			\hline
			{} & {} & {} & {} & \multicolumn{4}{@{}c@{}}{$\psi_{10}$}  & \multicolumn{4}{@{}c@{}}{$\psi_{11}$}  \\ 
			\cmidrule{5-8} \cmidrule{9-12}  
			n & $\rho$ & ($\gamma_{10}$, $\gamma_{01}$) & $\boldsymbol{\hat{\psi}}$ & \multicolumn{1}{c}{Bias} & \multicolumn{1}{c}{SE} & \multicolumn{1}{c}{RMSE} & \multicolumn{1}{c}{CR\%} & \multicolumn{1}{c}{Bias} & \multicolumn{1}{c}{SE} & \multicolumn{1}{c}{RMSE} & CR\%    \\ 
			\hline
			500 & 0.3 & (0.0, 0.0) & $\boldsymbol{\hat{\psi}^{v}}$ &  -0.005	& 0.722	& 0.722	& 93.6	& -0.014	& 0.607	& 0.607	& 94.2  \\ 
			{} & {} & (0.1, 0.1) & $\boldsymbol{\hat{\psi}^{n}}$ &  -0.175	& 0.161	& 0.238	& 85.2	& 0.175	& 0.124	& 0.214	& 77.2  \\
			{} & {} & {} & $\boldsymbol{\hat{\psi}^{mle}}$ &  0.016	& 0.224	& 0.225	& 95.4	& -0.016	& 0.186	& 0.187	& 93.0  \\
			{} & {} & (0.2, 0.2) & $\boldsymbol{\hat{\psi}^{n}}$ &  -0.285	& 0.152	& 0.323	& 65.8	& 0.288	& 0.114	& 0.310	& 46.4 \\
			{} & {} & {} & $\boldsymbol{\hat{\psi}^{mle}}$ &  0.007	& 0.267	& 0.267	& 94.8	& -0.014	& 0.222	& 0.222	& 93.2  \\ 
			{} & {} & (0.3, 0.3) & $\boldsymbol{\hat{\psi}^{n}}$ &  -0.362	& 0.148	& 0.391	& 48.4	& 0.366	& 0.108	& 0.382	& 24.0  \\
			{} & {} & {} & $\boldsymbol{\hat{\psi}^{mle}}$ &  0.009	& 0.307	& 0.307	& 92.6	& -0.020	& 0.249	& 0.250	& 94.6 \\ 
			\cmidrule{2-12} 
			{} & 0.5 & (0.0, 0.0) & $\boldsymbol{\hat{\psi}^{v}}$ &  0.002	& 0.262	& 0.262	& 95.2	& -0.014	& 0.210	& 0.210	& 94.4 \\ 
			{} & {} & (0.1, 0.1) & $\boldsymbol{\hat{\psi}^{n}}$ &  -0.177	& 0.161	& 0.239	& 89.4	& 0.171	& 0.124	& 0.211	& 87.6   \\
			{} & {} & {} & $\boldsymbol{\hat{\psi}^{mle}}$ &  -0.003	& 0.204	& 0.204	& 94.2	& -0.008	& 0.165	& 0.165	& 94.2  \\ 
			{} & {} & (0.2, 0.2) & $\boldsymbol{\hat{\psi}^{n}}$ & -0.285	& 0.152	& 0.323	& 75.8	& 0.288	& 0.114	& 0.310	& 63.8  \\
			{} & {} & {} & $\boldsymbol{\hat{\psi}^{mle}}$ & 0.001	& 0.226	& 0.226	& 95.0	& -0.010	& 0.184	& 0.184	& 94.2  \\  
			{} & {} & (0.3, 0.3) & $\boldsymbol{\hat{\psi}^{n}}$ & -0.367	& 0.147	& 0.395	& 58.4	& 0.370	& 0.107	& 0.385	& 37.8  \\
			{} & {} & {} & $\boldsymbol{\hat{\psi}^{mle}}$ &  0.005	& 0.245	& 0.245	& 94.8	& -0.014	& 0.198	& 0.198	& 95.0 \\ 
			\hline
			2000 & 0.3 & (0.0, 0.0) & $\boldsymbol{\hat{\psi}^{v}}$ &  0.008	& 0.158	& 0.158	& 94.2	& -0.007	& 0.126	& 0.126	& 95.0    \\
			{} & {} & (0.1, 0.1) & $\boldsymbol{\hat{\psi}^{n}}$ & -0.170	& 0.078	& 0.187	& 55.6	& 0.171	& 0.061	& 0.182	& 34.8    \\
			{} & {} & {} & $\boldsymbol{\hat{\psi}^{mle}}$ & 0.008	& 0.105	& 0.105	& 93.0	& -0.008	& 0.086	& 0.086	& 93.2  \\ 
			{} & {} & (0.2, 0.2) & $\boldsymbol{\hat{\psi}^{n}}$ & -0.287	& 0.074	& 0.296	& 11.0	& 0.288	& 0.055	& 0.293	& 1.0  \\
			{} & {} & {} & $\boldsymbol{\hat{\psi}^{mle}}$ &  0.004	& 0.123	& 0.123	& 93.8	& -0.007	& 0.100	& 0.100	& 93.4  \\ 
			{} & {} & (0.3, 0.3) & $\boldsymbol{\hat{\psi}^{n}}$ &  -0.374	& 0.072	& 0.381	& 1.6	& 0.376	& 0.052	& 0.380	& 0.0   \\
			{} & {} & {} & $\boldsymbol{\hat{\psi}^{mle}}$ & 0.000	& 0.139	& 0.139	& 92.6	& -0.005	& 0.113	& 0.113	& 93.8    \\ 
			\cmidrule{2-12}
			{} & 0.5 & (0.0, 0.0) & $\boldsymbol{\hat{\psi}^{v}}$ &  0.001	& 0.121	& 0.121	& 94.6	& -0.005	& 0.096	& 0.096	& 94.6   \\ 
			{} & {} & (0.1, 0.1) & $\boldsymbol{\hat{\psi}^{n}}$ &  -0.176	& 0.078	& 0.193	& 63.8	& 0.177	& 0.060	& 0.187	& 47.8 \\
			{} & {} & {} & $\boldsymbol{\hat{\psi}^{mle}}$ &  0.004	& 0.097	& 0.097	& 93.4	& -0.005	& 0.078	& 0.078	& 95.0  \\
			{} & {} & (0.2, 0.2) & $\boldsymbol{\hat{\psi}^{n}}$ & -0.287	& 0.074	& 0.296	& 24.2	& 0.288	& 0.055	& 0.293	& 5.0  \\
			{} & {} & {} & $\boldsymbol{\hat{\psi}^{mle}}$ & 0.002	& 0.106	& 0.106	& 94.2	& -0.005	& 0.085	& 0.085	& 93.6  \\ 
			{} & {} & (0.3, 0.3) & $\boldsymbol{\hat{\psi}^{n}}$ & -0.369	& 0.072	& 0.376	& 4.8	& 0.374	& 0.052	& 0.378	& 0.4  \\
			{} & {} & {} & $\boldsymbol{\hat{\psi}^{mle}}$ & 0.002	& 0.114	& 0.114	& 95.8	& -0.002	& 0.091	& 0.091	& 96.0  \\
			\midrule
		\end{tabular} 
	\end{table}

	\begin{sidewaystable}[hbt!]
		\setlength\tabcolsep{2pt} 
		\caption{Two-stage estimates of blip parameters $(\psi_{20}, \psi_{21}, \psi_{22}, \psi_{10}, \psi_{11})$}  
		\label{tab:two}
		\centering 
		\begin{adjustbox}{width=\textwidth}
		\begin{tabular}{ccrrccrrccrrccrrccrrccr}
			\hline
			\multicolumn{3}{@{}c@{}}{} &  \multicolumn{4}{@{}c@{}}{$\psi_{20}$} &  \multicolumn{4}{@{}c@{}}{$\psi_{21}$} &  \multicolumn{4}{c}{$\psi_{22}$} & \multicolumn{4}{@{}c@{}}{$\psi_{10}$}  & \multicolumn{4}{@{}c@{}}{$\psi_{11}$}  \\  
			\cmidrule{4-7} \cmidrule{8-11} \cmidrule{12-15}  \cmidrule{16-19}  \cmidrule{20-23} 
			$\rho$ & ($\gamma_{10}$, $\gamma_{01}$) & $\boldsymbol{\hat{\psi}}$ & \multicolumn{1}{c}{Bias} & \multicolumn{1}{c}{SE} & \multicolumn{1}{c}{RMSE} & \multicolumn{1}{c}{CR\%} & \multicolumn{1}{c}{Bias} & \multicolumn{1}{c}{SE} & \multicolumn{1}{c}{RMSE} & \multicolumn{1}{c}{CR\%} & \multicolumn{1}{c}{Bias} & \multicolumn{1}{c}{SE} & \multicolumn{1}{c}{RMSE} & \multicolumn{1}{c}{CR\%} & \multicolumn{1}{c}{Bias} & \multicolumn{1}{c}{SE} & \multicolumn{1}{c}{RMSE} & \multicolumn{1}{c}{CR\%} & \multicolumn{1}{c}{Bias} & \multicolumn{1}{c}{SE} & \multicolumn{1}{c}{RMSE} &  \multicolumn{1}{c}{CR\%}  \\  
			\hline
			0.3 & (0.0, 0.0) & $\boldsymbol{\hat{\psi}^{v}}$ &  0.005	& 0.143	& 0.143	& 93.2	& 0.004	& 0.111	& 0.111	& 95.4	& 0.001	& 0.139	& 0.139	& 93.8 & -0.007	& 0.053	& 0.053	& 93.6	& -0.004	& 0.047	& 0.047	& 92.8    \\ 
			{} & (0.1, 0.1) & $\boldsymbol{\hat{\psi}^{n}}$ & -0.076	& 0.066	& 0.101	& 86.2	& -0.169	& 0.052	& 0.177	& 25.4	& -0.171	& 0.063	& 0.182	& 39.4 & 0.127	& 0.019	& 0.128	& 76.6	& -0.012	& 0.017	& 0.021	& 94.0  \\
			{} & {} & $\boldsymbol{\hat{\psi}^{mle}}$ & 0.003	& 0.090	& 0.090	& 92.4	& 0.000	& 0.073	& 0.073	& 94.4	& 0.007	& 0.089	& 0.089	& 93.2 & -0.001	& 0.029	& 0.029	& 92.8	& -0.009	& 0.026	& 0.028	& 93.2   \\
			{} & (0.2, 0.2) & $\boldsymbol{\hat{\psi}^{n}}$ & -0.132	& 0.062	& 0.146	& 54.8	& -0.280	& 0.049	& 0.284	& 0.0	& -0.284	& 0.058	& 0.290	& 2.0 & 0.211	& 0.012	& 0.211	& 41.4	& -0.011	& 0.011	& 0.016	& 94.0     \\
			{} & {} & $\boldsymbol{\hat{\psi}^{mle}}$ &  0.005	& 0.107	& 0.107	& 94.8	& 0.004	& 0.086	& 0.086	& 95.4	& 0.009	& 0.105	& 0.105	& 91.0 & 0.003	& 0.029	& 0.029	& 93.4	& -0.001	& 0.026	& 0.026	& 94.6   \\
			{} & (0.3, 0.3) & $\boldsymbol{\hat{\psi}^{n}}$ & -0.174	& 0.061	& 0.184	& 34.4	& -0.364	& 0.047	& 0.367	& 0.0	& -0.368	& 0.055	& 0.372	& 0.0 & 0.271	& 0.008	& 0.271	& 11.6	& -0.016	& 0.007	& 0.017	& 94.4   \\
			{} & {} & $\boldsymbol{\hat{\psi}^{mle}}$ &  0.003	& 0.124	& 0.124	& 93.8	& 0.004	& 0.098	& 0.098	& 95.4	& 0.002	& 0.120	& 0.120	& 94.6 & -0.006	& 0.029	& 0.030	& 94.2	& -0.004	& 0.026	& 0.026	& 94.4  \\ 
			\hline
			0.5 & (0.0, 0.0) & $\boldsymbol{\hat{\psi}^{v}}$ &  0.002	& 0.107	& 0.107	& 94.0	& 0.004	& 0.085	& 0.085	& 93.8	& 0.013	& 0.103	& 0.104	& 92.4 & -0.003	& 0.041	& 0.041	& 93.8	& 0.004	& 0.037	& 0.037	& 94.6   \\ 
			{} & (0.1, 0.1) & $\boldsymbol{\hat{\psi}^{n}}$ & -0.077	& 0.066	& 0.101	& 86.0	& -0.170	& 0.052	& 0.178	& 39.4	& -0.171	& 0.063	& 0.182	& 52.8 & 0.127	& 0.019	& 0.128	& 81.2	& -0.005	& 0.017	& 0.018	& 95.0  \\
			{} & {} & $\boldsymbol{\hat{\psi}^{mle}}$ & 0.001	& 0.085	& 0.085	& 93.8	& 0.003	& 0.068	& 0.068	& 94.6	& 0.008	& 0.082	& 0.082	& 93.8 &  -0.003	& 0.029	& 0.029	& 94.2	& 0.004	& 0.026	& 0.026	& 93.8     \\
			{} & (0.2, 0.2) & $\boldsymbol{\hat{\psi}^{n}}$ & -0.134	& 0.063	& 0.148	& 67.4	& -0.282	& 0.048	& 0.286	& 3.6	& -0.283	& 0.058	& 0.289	& 8.8 & 0.210	& 0.012	& 0.210	& 54.6	& -0.008	& 0.011	& 0.014	& 93.8   \\
			{} & {} & $\boldsymbol{\hat{\psi}^{mle}}$ &  0.000	& 0.094	& 0.094	& 94.6	& 0.003	& 0.075	& 0.075	& 94.8	& 0.011	& 0.090	& 0.091	& 93.0 & -0.003	& 0.029	& 0.029	& 94.4	& 0.005	& 0.026	& 0.026	& 95.0  \\ 
			{} & (0.3, 0.3) & $\boldsymbol{\hat{\psi}^{n}}$ & -0.176	& 0.061	& 0.186	& 46.4	& -0.366	& 0.046	& 0.369	& 0.0	& -0.365	& 0.055	& 0.369	& 0.2 & 0.279	& 0.008	& 0.279	& 25.6	& -0.014	& 0.007	& 0.016	& 93.2   \\
			{} & {} & $\boldsymbol{\hat{\psi}^{mle}}$ &  0.000	& 0.100	& 0.100	& 93.8	& 0.003	& 0.080	& 0.080	& 93.4	& 0.007	& 0.097	& 0.097	& 95.6 & -0.002	& 0.029	& 0.029	& 91.6	& 0.003	& 0.026	& 0.026	& 94.0  \\ 
			\hline
		\end{tabular}
	\end{adjustbox}
	\end{sidewaystable}

	\begin{sidewaystable}[ht!]
		\centering
		\caption{}
		\label{tab:prediction}
		\begin{adjustbox}{width=\textwidth}
		\begin{tabular*}{\textwidth}{@{\extracolsep{\fill}}ccccccccccc@{\extracolsep{\fill}}}
			\multicolumn{11}{c}{Panel A: Prediction accuracy of optimal DTR} \\
			\hline
			{} & {} & \multicolumn{3}{c}{Stage 2 (\%)} & \multicolumn{3}{c}{Stage 1 (\%)} & \multicolumn{3}{c}{Stage 2 \& Stage 1 (\%)}  \\ 
			\cmidrule{3-5} \cmidrule{6-8} \cmidrule{9-11} 
			$\rho$ & ($\gamma_{10}$, $\gamma_{01}$)  & v & n & mle & v & n & mle & v & n & mle  \\ 
			\hline
			0.3 &  (0.1, 0.1) &  91.8	& 97.8	& 98.6	& 93.6	& 97.8	& 98.6	& 88.6	& 96.7	& 97.9	  \\
			{} &  (0.2, 0.2)  &  91.7	& 92.1	& 95.8	& 92.7	& 93.3	& 97.0	& 88.1	& 88.8	& 94.4   \\
			{} &  (0.3, 0.3)   &  91.9	& 84.3	& 94.7	& 92.9	& 86.9	& 96.4	& 88.4	& 77.6	& 93.0    \\ 
			\hline
			0.5 & (0.1, 0.1)  & 95.6	& 97.3	& 98.8	& 97.1	& 98.1	& 99.5	& 94.2	& 96.3	& 98.6    \\
			{} & (0.2, 0.2)  &  96.3	& 91.2	& 98.0	& 97.3	& 92.7	& 98.8	& 94.9	& 87.5	& 97.4     \\
			{} &  (0.3, 0.3)  &  96.7	& 81.6	& 97.6	& 97.3	& 81.4	& 98.1	& 95.4	& 72.2	& 96.6      \\
			\hline
			\\
			\multicolumn{11}{c}{Panel B: Predicted error rates, sensitivity, and specificity of the outcome} \\
			\hline
			{} & {} & \multicolumn{3}{c}{Error Rates (\%)} & \multicolumn{3}{c}{Sensitivity (\%)} & \multicolumn{3}{c}{Specificity (\%)}  \\ 
			\cmidrule{3-5} \cmidrule{6-8} \cmidrule{9-11} 
			$\rho$ & ($\gamma_{10}$, $\gamma_{01}$)  & v & n & mle  & v & n & mle  & v & n & mle  \\ 
			\hline
			0.3 &  (0.1, 0.1) & 5.2	& 3.8	& 3.3	& 96.0	& 97.2	& 97.4	& 92.9	& 94.7	& 95.6    \\
			{} &  (0.2, 0.2)  & 5.5	& 5.8	& 4.2	& 95.7	& 95.9	& 96.6	& 92.7	& 91.5	& 94.7 	 \\
			{} &  (0.3, 0.3)   & 5.4	& 8.5	& 4.6	& 95.8	& 94.7	& 96.4	& 92.8	& 86.5	& 94.0    \\ 
			\hline
			0.5 & (0.1, 0.1)  &  4.0	& 3.8	& 3.0	& 96.7	& 97.0	& 97.5	& 95.0	& 94.9	& 96.1 	  \\
			{} & (0.2, 0.2)  &  3.9	& 5.9	& 3.4	& 96.9	& 95.3	& 97.1	& 95.0	& 92.2	& 95.8	  \\
			{} &  (0.3, 0.3)  &  4.0	& 9.3	& 3.8	& 96.7	& 93.9	& 96.9	& 94.8	& 85.8	& 95.3 	 \\
			\hline
		\end{tabular*}
	\end{adjustbox}
	\begin{tablenotes}
		\item[1] v: validation estimator, n: naive estimator, mle: MLE estimator
	\end{tablenotes}
	\end{sidewaystable}

	\begin{sidewaystable}[ht!]
		\centering
		\caption{}
		\label{tab:data_analysis}
		\begin{adjustbox}{width=\textwidth}
		\begin{tabular*}{\textwidth}{@{\extracolsep{\fill}}crrrrrrrrrr@{\extracolsep{\fill}}}
			\multicolumn{11}{c}{Panel A: Sensitivity analysis results of the NHEFS data for the blip estimators}\label{tab:NHEFS}  \\
			\hline
			{} & {} & \multicolumn{3}{@{}c@{}}{A}  & \multicolumn{3}{@{}c@{}}{A*Diabetes} & \multicolumn{3}{@{}c@{}}{A*SmokeIntensity} \\  
			\cmidrule{3-5}\cmidrule{6-8} \cmidrule{9-11} 
			Method &  \multicolumn{1}{c}{$\gamma_{10}$} &  \multicolumn{1}{c}{Est} & \multicolumn{1}{c}{SE} & \multicolumn{1}{c}{95\%CI} & \multicolumn{1}{c}{Est} & \multicolumn{1}{c}{SE} & \multicolumn{1}{c}{95\%CI} & \multicolumn{1}{c}{Est} & \multicolumn{1}{c}{SE} & \multicolumn{1}{c}{95\%CI}  \\ 
			\hline
			Naive & {} & -0.148	& 0.091	& (-0.300,	 0.027)	& 0.130	& 0.066 & (0.005,	0.257)	& 0.075 & 0.070	& (-0.037,	0.236) \\
			MLE & 5\% & -0.179	& 0.113 & (-0.376,	0.034)	& 0.149 & 0.079	& (-0.017,	 0.304)	& 0.121	& 0.097	& (-0.023,	0.343) \\
			{}  & 7.5\% &  -0.201	& 0.129	& (-0.433,	0.039)	& 0.160	& 0.089	& (-0.025,	 0.342)	& 0.148	& 0.114	& (-0.011,	0.399) \\
			{}  & 10\% &  -0.231	& 0.151	& (-0.514,	0.047) & 	0.173	& 0.101	& (-0.034,	0.374)	& 0.179	& 0.136	& (-0.005,	0.470) \\
			{}  & 12.5\% &  -0.270	& 0.186	& (-0.618,	0.077)	& 0.187	& 0.117	& (-0.054,	0.409)	& 0.213	& 0.174	& (-0.048,	0.565) \\ 
			\hline
			\\
			\multicolumn{11}{c}{Panel B: Sensitivity analysis results of the smoking cessation data for the blip estimators}\label{tab:smoking} \\
			\hline
			{} & {} & \multicolumn{3}{@{}c@{}}{A}  & \multicolumn{3}{@{}c@{}}{A*HTN} & \multicolumn{3}{@{}c@{}}{A*YrsSmoke} \\  
			\cmidrule{3-5}\cmidrule{6-8} \cmidrule{9-11} 
			Method & \multicolumn{1}{c}{$\gamma_{10}$} & \multicolumn{1}{c}{Est} & \multicolumn{1}{c}{SE} & \multicolumn{1}{c}{95\%CI} & \multicolumn{1}{c}{Est} & \multicolumn{1}{c}{SE} & \multicolumn{1}{c}{95\%CI} & \multicolumn{1}{c}{Est} & \multicolumn{1}{c}{SE} & \multicolumn{1}{c}{95\%CI}  \\ 
			\hline
			Naive &  {} &  0.628	& 0.450	& (-0.253,	1.509)	& -0.244	& 0.628	& (-1.474,	0.986)	& -0.162	& 0.404	& (-0.954,	0.629)     \\ 
			MLE & 2.5\% & 0.741	& 1.287	& (-1.781,	3.263)	& -0.239	& 1.021	& (-2.241,	1.762)	& -0.185	& 1.145	& (-2.429,	2.059)    \\ 
			{} & 5\% & 2.487	& 1.329	& (-0.117,	5.092)	& -0.536	& 1.375	& (-3.230,	2.159)	& 1.051	& 1.396	& (-1.685,	3.787)   \\ 
			{} & 7.5\% & 2.811	& 1.257	& (0.348,	5.274)	& -0.651	& 1.606	& (-3.798,	2.497)	& 1.214	& 1.423	& (-1.575,	4.003)    \\ 
			{} & 8.5\% & 2.903	& 1.154	& (0.641,	5.164)	& -0.692	& 1.614	& (-3.854,	2.471)	& 1.254	& 1.454	& (-1.596,	4.105)  \\ 
			\hline
		\end{tabular*}
	\end{adjustbox}
		\begin{tablenotes}
			\item[1] Est: estimates, SE: standard error, CI: confidence interval
		\end{tablenotes}
	\end{sidewaystable}
	
\end{appendices}

\end{document}